\documentclass{sig-alternate}

\usepackage[
  pass,% keep layout unchanged
]{geometry}

\usepackage{titlesec}
\usepackage{subcaption}
\usepackage[frozencache,cachedir=.]{minted}
\begin{document}

% Copyright
\setcopyright{waclicense}

%% DOI
%\doi{10.475/123_4}
%
%% ISBN
%\isbn{123-4567-24-567/08/06}
%
%%Conference
\conferenceinfo{Web Audio Conference WAC-2024,}{March 15--17, 2024, West Lafayette, IN, USA.}
\CopyrightYear{2024} % Allows default copyright year (20XX) to be over-ridden - IF NEED BE.
%\crdata{0-12345-67-8/90/01}  % Allows default copyright data (0-89791-88-6/97/05) to be over-ridden - IF NEED BE.
% --- End of Author Metadata ---

\title{Creating Aesthetic Sonifications on the Web with SIREN}
%\subtitle{[Extended Abstract]
%\titlenote{A full version of this paper is available as
%\textit{Author's Guide to Preparing ACM SIG Proceedings Using
%\LaTeX$2_\epsilon$\ and BibTeX} at

\numberofauthors{3} %  in this sample file, there are a *total*
\author{
% 1st. author
\alignauthor
       \name{Tristan Peng}
       \affaddr{Center for Computer Research in Music and Acoustics\\Stanford University, Stanford, CA}
       \email{pengt@ccrma.stanford.edu}
% 2nd. author
\alignauthor
       \name{Hongchan Choi}
       \affaddr{Center for Computer Research in Music and Acoustics\\Stanford University, Stanford, CA}
       \email{hongchan@ccrma.stanford.edu}
% 3rd. author
\alignauthor
       \name{Jonathan Berger}
       \affaddr{Center for Computer Research in Music and Acoustics\\Stanford University, Stanford, CA}
       \email{brg@ccrma.stanford.edu}
% \and  % use '\and' if you need 'another row' of author names
% % 4th. author
% \alignauthor
%        \name{Lawrence P. Leipuner}
%        \affaddr{Brookhaven Laboratories\\ Brookhaven National Lab}
%        \email{lleipuner@researchlabs.org}
% % 5th. author
% \alignauthor
%        \name{Sean Fogarty}
%        \affaddr{NASA Ames Research Center\\ Moffett Field}
%        \email{fogartys@amesres.org}
% % 6th. author
% \alignauthor
%        \name{Charles Palmer}
%        \affaddr{Palmer Research Laboratories\\ 8600 Datapoint Drive}
%        \email{cpalmer@prl.com}
}
% There's nothing stopping you putting the seventh, eighth, etc.
% author on the opening page (as the 'third row') but we ask,
% for aesthetic reasons that you place these 'additional authors'
% in the \additional authors block, viz.
% \additionalauthors{Additional authors: John Smith (The Th{\o}rv{\"a}ld Group,
% email: {\texttt{jsmith@affiliation.org}}) and Julius P.~Kumquat
% (The Kumquat Consortium, email: {\texttt{jpkumquat@consortium.net}}).}
\date{\today}
% Just remember to make sure that the TOTAL number of authors
% is the number that will appear on the first page PLUS the
% number that will appear in the \additionalauthors section.

\maketitle
\begin{sloppypar}
\begin{abstract}
SIREN is a flexible, extensible, and customizable web-based general-purpose interface for auditory data display (sonification).
Designed as a digital audio workstation for sonification, synthesizers written in JavaScript using the Web Audio API facilitate intuitive mapping of data to auditory parameters for a wide range of purposes.

This paper explores the breadth of sound synthesis techniques supported by SIREN, and details the structure and definition of a SIREN synthesizer module. The paper proposes further development that will increase SIREN's utility.
\end{abstract}


%% End generated code
%%
%
%%
%%  Use this command to print the description
%%
%\printccsdesc
%
%% We no longer use \terms command
%%\terms{Theory}
%
%\keywords{ACM proceedings, \LaTeX, text tagging}

\section{Introduction}
Over a quarter of a century ago, Myers et al stated,  “It remains an important challenge to find ways to achieve the very desirable outcome of systems with both a low threshold and a high ceiling at the same time" \cite{Myers_Hudson_Pausch_2000}. Even now, this remains a challenge to find an optimal balance between increasing the utility of a system while remaining accessible for anyone to use. In the realm of sonification applications, many unique iterations have been made, landing on myriad locations along the spectrum. As detailed below in section \ref{sec:related work}, some sonification interfaces, especially those for a specific application, provide a more opinionated set of controls for the user, while others try to create a more universal design, allowing for general-purpose sonification. Sonification Interface for REmapping Nature (SIREN) attempts to optimize the threshold-ceiling tension. While in some aspects, SIREN can have a higher threshold than other sonification interfaces, SIREN balances this by expediting other parts of the sonification creation process through an accessible user interface and intuitive workflow.

SIREN uses parameter mapping as its main paradigm for sonification. Parameter mapping sonification is a powerful and multidimensional method for data sonification \cite{Hermann_Hunt_Neuhoff_2011, Barrass_Kramer_1999}. Previous papers on SIREN discuss SIREN's development from version 0.1 to 0.3, and have not presented on the breadth of sound synthesis methods that SIREN supports \cite{Peng_Choi_2021, Peng_Choi_Berger_2023}. In this paper, we discuss some of the improvements in SIREN's version 0.4, as well as heavily focus on the range of sound synthesis methods that SIREN supports that advocate for SIREN as a truly general-purpose data sonification tool. Future work will discuss the possibilities of expanding further to develop and support more sound synthesis methods.
\section{Related Work}
\label{sec:related work}
The extensibility of sonification interfaces is a crucial aspect of its design in order for users to customize their sonification to their specifications. There have been a plethora of sonification applications developed throughout the years, and many of them provide an opinionated set of parameters that allow for accessible sonification for a particular sonification paradigm.
\subsection{MUsical Sonification Environment} % 1997
MUsical Sonification Environment is one of the earlier iterations of an interface for data sonification. Because the application was developed more to explore the musification of data, more emphasis was placed on parameters that would sound musical. Thus, the mappings from data to auditory parameters include pre-made instrumental and voice timbres, rhythms emulating different musical styles in the Western classical music tradition, tempo, gain, pitch, and harmony (where data can be mapped on a scale of consonance to dissonance) \cite{Lodha_Beahan_Heppe_Joseph_Zane-Ulman_1997}.
\subsection{MUSical Audio transfer function Realtime Toolkit} % 2002
MUSical Audio transfer function Realtime Toolkit (MUSART) presents an innovative approach to formulating parameter mapping sonification. MUSART's team developed the audio transfer function, which sketches a mapping schema where data can be applied to up to 10 different auditory parameters, including pitch, register (related to pitch but defines the range of pitch), timbre (selected out of nine options), thickness (layering chord tones), duration, silence, gain, pan, beats (tremolo), and harmony (consonance and dissonance) \cite{Joseph_Lodha_2002}.
\subsection{Sonification Application and Research Toolbox} % 2002
The Sonification Application and Research Toolbox (SonART), developed on top of the Synthesis ToolKit (STK), allows users to create sonifications by controlling all of the parameters that STK instruments enable \cite{Cook_Scavone_1999}. Through the parameter engine, SonART enables parameter mapping of a variety of parameters through a variety of sound synthesis methods. This symbiotic relationship allows SonART to be very flexible and customizable in terms of the parameter mapping controls it affords its users \cite{Ben-Tal_Berger_Cook_Daniels_Scavone_2002}.
\subsection{Sonification Sandbox} % 2003
Sonification Sandbox has been advertised as the multi-platform general purpose sonification tool, and it is quite powerful as a sonification tool. Beyond pitch, timbre, gain, and pan parameterization for every data set, Sonification Sandbox also includes contextual sonic notifications that help contextualize the sonification within the entire dataset \cite{Walker_Cothran_2003}. An update to Sonification Sandbox in 2007 presented a more cohesive auditory graph data model for representing sonifications, but expanded less on the controllable parameters that Sonification Sandbox provides for users \cite{Davison_Walker_2007}.
\subsection{Interactive Sonification Toolkit} % 2004
Interactive Sonification Toolkit identified various categories that ensure perceptually differentiable sonifications. Among these include parameter mapping sonification onto auditory parameters such as pitch, gain, duration, timbre, pan. The toolkit also supports audification, and results have demonstrated a myriad of sonification types that can be produced with these interactions \cite{Pauletto_Hunt_2004}.
\subsection{Sonification Integrative Flexible Toolkit}
Like many other sonification applications, Sonification Integrative Flexible Toolkit provides a set list of auditory parameters that the user can parameterize. These include pitch, gain, timbre, duration, and pan \cite{Bruce_Palmer_2005}.
\subsection{xSonify} % 2006
Created as an accessible tool for visually-impaired researchers in space science, xSonify is specifically tailored to the application of sonifying space science data. As an opinionated tool, xSonify has a limited set of customizable parameters: pitch, gain, and rhythm \cite{Candey_Schertenleib_Diaz-Merced_2006}. A follow-up study of xSonify in 2019 has proposed a more accessible and modern interface to further aid visually-impaired researchers, though the paper still limits the user's customizability of parameters to the same three \cite{Garcia_Diaz-Merced_Casado_Cancio_2019}.
\subsection{Sonifyer} % 2008
Audification is Sonifyer's main avenue for sonification, due to its direct applications in sonifying electroencephalogram and earthquake data, but it also has capability for parameter mapping sonification in the form of frequency modulation synthesis \cite{Dombois_2008}.
\subsection{Rotator} % 2016
Rotator, developed as a tool for distribution of data across the senses, focuses more on the audiovisual display of data. As such, it incorporates six synthesizer types into its sonification capabilities, including noise, envelope, clicks, oscillator, beats, and direct \cite{Cherston_Paradiso_2017}.
\subsection{Sonification Workstation} % 2019
Sonification Workstation presents a unique interfacing for creating sonifications. Inspired by visual programming languages like Max/MSP and Pure Data, Sonification Workstation includes components that can be patched together, with settings that can be controlled by data. These include the oscillator, amplitude and frequency modulation, audification, pan, envelope, gain, noise, and equalizer components \cite{Phillips_Cabrera_2019}.
\subsection{WebAudioXML Sonification Toolkit} % 2021
WebAudioXML was created as an XML syntax alternative to the Web Audio API without having to learn the intricacies of JavaScript \cite{Lindetorp_Falkenberg_2020}. Based upon this technology, WebAudioXML Sonification Toolkit was subsequently developed. The sonification interface contains a permutation of parameters based on the timbre, including pitch, brightness, gain, filter, playback rate, thinness, and trigger rate \cite{Lindetorp_Falkenberg_2021}.
\subsection{Highcharts Sonification Studio} % 2021
The successor of Sonification Sandbox, Highcharts Sonification Studio is a web-based, open-source general purpose sonification application. As Highcharts Sonification Studio was designed as a modern web port of Sonification Sandbox, many of the features that Sonification Sandbox afforded users are also paralleled in Highcharts Sonification Studio \cite{Cantrell_Walker_Moseng_2021}. This includes the pitch, pan, gain, and timbre controls from Sonification Sandbox, but expands upon it with lowpass and highpass filter and tremolo depth and speed parameterizations.
\section{SIREN}
SIREN has had many previous iterations, and every iteration has grown in terms of user customizability and extensibility \cite{Peng_Choi_2021, Peng_Choi_Berger_2023}. The version presented in this paper, version 0.4, allows for a multitude of sound synthesis methods while retaining a clean, accessible interface. First, we will briefly discuss SIREN's design and architecture, workflow, and finally detail the intricacies of creating aesthetic sonifications through SIREN's sound synthesis paradigm.
\subsection{Design}
\begin{figure*}[!h]
    \centering
    \fbox{\includegraphics[width=0.7\textwidth]{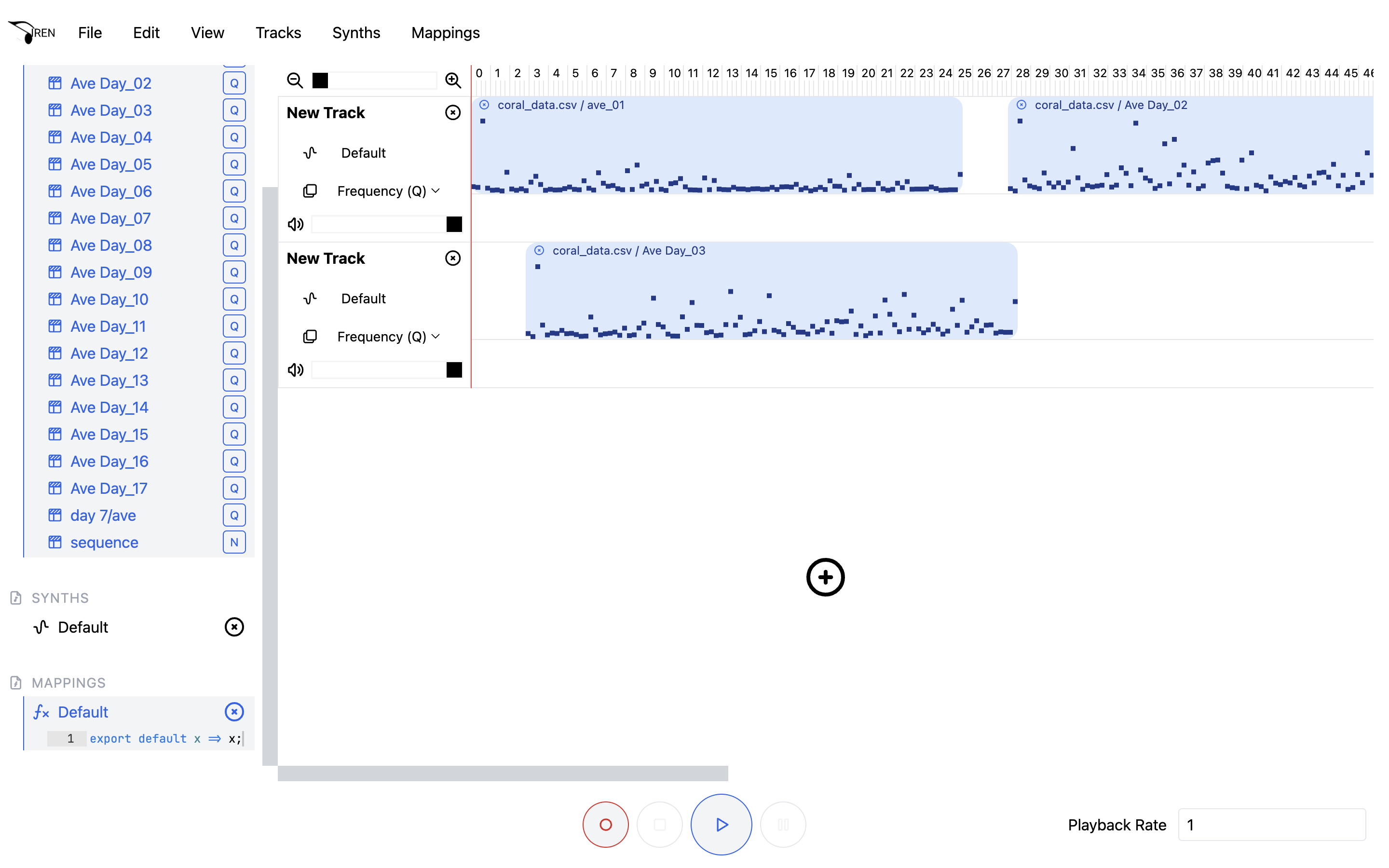}}
    \caption{SIREN's web interface.}
    \label{fig:interface}
\end{figure*}
SIREN's interface is heavily inspired by those of digital audio workstations. Different than other sonification applications, this layout allows users to visualize multidimensional and polyphonic sonifications easily.

The main organizational unit of sonification in SIREN is the \textbf{track}, visualized by the horizontal boxes headed by in this case, the titles ``Track 1'' and ``Track 2'' stretching all the way to the right seen in figure \ref{fig:interface}. In the case of figure \ref{fig:interface}, there are two tracks currently created. A track serves two important purposes: first, each track encapsulates one parameter mapped synthesizer; second, each track contains \textbf{regions}, visualized by the blue areas within the track bodies in figure \ref{fig:interface}, a structure that holds one column of the dataset. In figure \ref{fig:interface}, there are three regions, two in the track titled ``Track 1'' and one in ``Track 2.'' When the play button below all of the tracks is triggered, the red seeker will sweep from left to right, applying the values in the region to the parameter that it controls as they intersect the seeker.
\subsection{Architecture}
SIREN uses many modern tools and frameworks as part of its technology stack. The main framework that SIREN uses is SvelteKit, a new framework for building interactive web applications, and consequently uses Vite as a development environment. The codebase is mainly written in TypeScript.
\subsection{Workflow}
Creating sonifications in SIREN consists of four main steps, detailed below:
\begin{enumerate}
    \item \textbf{Adding datasets:} First, the user should load the dataset as a CSV file into SIREN by selecting ``File'' > ``Import CSV.'' This will make an entry appear under the ``Data'' section of the left sidebar. In figure \ref{fig:interface}, the default preloaded dataset ``coral\_data.csv'' is loaded and selected, revealing all of the columns that SIREN is able to parse from the dataset, along with an indicator ``N'' or ``Q'' representing the data type of the column.
    \item \textbf{Adding synthesizers:} By selecting ``Synths'' > ``Add Synth,'' the user can then load a parameter-mapped synthesizer, of which the specification is detailed in section \ref{ssec:synthesis}.
    \item \textbf{Creating tracks and regions:} The user can first create tracks by double-clicking on the plus button in the middle of the interface or by selecting ``Tracks'' > ``Add Track.'' Then, by dragging and dropping any column onto the track body, the user can create regions. The parameters that the regions automate can be selected using the selection menu in the track header, and the synthesizer used for that track can be changed by dragging the synthesizer from the ``Synths'' section in the left sidebar onto the synth name in the track header.
    \item \textbf{Applying mappings:} Since data may not conform exactly to the bounds of the parameters, users can create mappings using ``Mappings'' > ``Add Mapping,'' from which they can write code to transform the data of a region by dragging the mapping from the left sidebar onto a region.
\end{enumerate}
After these four tasks, the user can then play the sonification directly using the playback controls on the bottom of the screen (as seen in figure \ref{fig:interface}) or exporting it through the toolbar by clicking ``File > Export .wav.''
\subsection{Sound Synthesis Paradigm}
\label{ssec:synthesis}
SIREN's sound synthesis paradigm lands on the higher complexity end of the threshold-ceiling spectrum. However, many efforts have been made in the design of SIREN's sound synthesis paradigm to make it as accessible as possible for users to create their own sounds for their purposes. As a web-based interface for sonification, SIREN currently accepts JavaScript modules as input for the synthesizer. Following the conventions that SIREN sets, users writing in JavaScript can synthesize their own sounds for sonification, allowing for a virtually unlimited set of different parameters that users can customize.

First, the structure of the synthesizer will be defined, and then the structure will be applied to various examples of synthesizers that have been tested with SIREN. These example synthesizers will show a small section of the vast breadth of different sound synthesis methods that SIREN is able to support for sonification.
\subsubsection{Synthesizer Structure}
SIREN enforces a simple schema for the module exports, allowing for maximal flexibility for users to structure their code. There are only two exports that SIREN requires of its synthesizer modules, the default export and a named export called \verb|parameters|. Beyond these two, SIREN also accepts an optional named export \verb|worklets| that allow users to create inline \verb|AudioWorklet|s to enable even more powerful sound synthesis techniques.

\begin{listing}[!h]
    \centering
\begin{minted}{ts}
interface Parameters {
  timbral: { [key: string]: 'nominal' | 'quantitative' };
  time: string[];
}
\end{minted}
    \caption{The definition of the \texttt{parameters} structure.}
    \label{list:parameters}
\end{listing}
\textbf{The \texttt{parameters} Export}
Firstly, the user should define a named export \verb|parameters| that defines name and data type of the various parameters that the synthesizer should expose to the user. The interface that defines this structure can be seen in listing \ref{list:parameters}. There are two types of parameters: timbral and temporal. Timbral parameters define parameters that affect the time-independent components of the sound, such as frequency, gain, or pan as an example. On the other hand, temporal parameters, keyed as \verb|time|, are time-dependent components, such as duration or grain length as an example. While temporal parameters should all have the same data type, that being a duration, timbral parameters can be nominal or quantitative, data types defined by Stanley Stevens \cite{Stevens_1946}. Ordinal data types are excluded as they can be created as a more restrictive version of a quantitative parameter. Thus, timbral parameters also must define, in addition to their name, what data type it takes on. This has a significant impact on the sonified result, since a quantitative dataset will sound different when sonified quantitatively versus nominally.

\begin{listing}[!h]
    \centering
\begin{minted}{ts}
(context: AudioContext, destination: AudioNode) => {
  updates: Map<string[], (...parameters: any[]) => any>;
  start: () => any;
  stop: () => any;
};
\end{minted}
    \caption{The definition of the synthesizer structure.}
    \label{list:default}
\end{listing}
\textbf{The Default Export}
The main synthesizer is defined in the default export. In order to write a compatible synthesizer module, the default export has a function signature represented in listing \ref{list:default}. To maintain maximum flexibility for the user, this function can be \verb|async| should the user need to do asynchronous tasks within the synthesizer such as fetching buffers for a granular synthesizer module.

The default exported function accepts two parameters \verb|context|, the \verb|AudioContext| upon which the user can create their audio graph using the Web Audio API, and \verb|destination|, the final audio node that the graph should route to. Once the user has set up the synthesizer and has properly routed audio on the \verb|context|, the user returns an object with three crucial pieces of information. The \verb|updates| property specifies how the parameter interacts with the synthesizer and changes the sonic output. As seen in the function definition, \verb|updates| is a \verb|Map| object that stores a key of an array of strings with a function value. The key defines the names of the properties that the particular update function manipulates, while the value defines the actual update function. The arguments of the function are those parameters defined by the map key, and the function body can use these arguments to alter different audio parameters to change the final output. Finally, the two other properties, \verb|start| and \verb|stop|, give space for the user to define actions or potentially \verb|AudioNode|s that need to be run when the sonification starts and stops.

\textbf{The \texttt{worklets} Export}
Since the inception of \verb|AudioWorklet|s, it has always been one of the most important audio nodes in the Web Audio API. However, since worklet code runs separately from the main JavaScript runtime, they have always been needed to be written in a separate file. However in SIREN's synthesizer module, new in version 0.4 of SIREN, users can write inline \verb|AudioWorklet|s that can be loaded and used as usual, and the optional named \verb|worklets| export is where the user should write these. \verb|worklets| should export an object, where the keys are both the names and module URLs of the \verb|AudioWorklet| and the values are the actual definition of the \verb|AudioWorklet|. When writing inline worklets for SIREN synthesizer modules, there is no need for \verb|registerProcessor|, as SIREN handles that when it reads the synthesizer.

\subsubsection{Working Example: Frequency Modulation Synthesis in SIREN}
% \begin{figure}[!h]
%     \centering
%     \begin{subfigure}[!h]{0.45\textwidth}
%         \centering
%         % \includegraphics[width=\textwidth]{assets/fm-parameters.png}
%         \inputminted[fontsize=\small]{js}{code/fm-parameters.ts}
%         \caption{FM synthesizer parameters.}
%         \label{list:fm parameters}
%     \end{subfigure}
% \end{figure}
% \begin{figure}[!h]
%     \centering
%     \ContinuedFloat
%     \begin{subfigure}[!h]{0.45\textwidth}
%         \centering
%         \includegraphics[width=\textwidth]{assets/fm-default.png}
%         \caption{The main synthesizer code for the FM synthesizer.}
%         \label{list:fm default}
%     \end{subfigure}
%     \caption{Basic FM synthesizer for SIREN.}
%     \label{list:fm}
% \end{figure}
\begin{listing}[!h]
    \centering
\begin{minted}{js}
export const parameters = {
  timbral: {
    p4: 'quantitative',
    p5: 'quantitative',
    p6: 'quantitative',
    p7: 'quantitative',
    p8: 'quantitative'
  },
  time: ['p3']
};

export default (context, destination) => {
  const car = new OscillatorNode(context);
  const eg1 = new ConstantSourceNode(context);
  const amp = new GainNode(context);

  const eg2 = new ConstantSourceNode(context);
  const mod = new OscillatorNode(context);
  const modAmp = new GainNode(context);

  eg2.connect(modAmp.gain);
  mod.connect(modAmp).connect(car.frequency);
  eg1.connect(amp.gain);
  car.connect(amp).connect(destination);

  return ({
    updates: new Map()
      .set(['p4', 'p3'], (p4 = 1, p3 = 0) =>
        eg1.offset.setValueAtTime(p4, p3))
      .set(['p5', 'p3'], (p5 = 440, p3 = 0) =>
        car.frequency.setValueAtTime(p5, p3))
      .set(['p6', 'p3'], (p6 = 440, p3 = 0) =>
        mod.frequency.setValueAtTime(p6, p3))
      .set(['p6', 'p7', 'p3'], (p6 = 440, p7 = 0, p3 = 0) =>
        modAmp.gain.setValueAtTime(p7 * p6, p3))
      .set(['p6', 'p7', 'p8', 'p3'],
        (p6 = 440, p7 = 0, p8 = 5, p3 = 0) =>
          eg2.offset.setValueAtTime((p8 - p7) * p6, p3)),
    start: () => {
      eg1.start();
      eg2.start();
      mod.start();
      car.start();
    },
    stop: () => {
      eg1.stop();
      eg2.stop();
      mod.stop();
      car.stop();
    }
  });
};
\end{minted}
    \caption{Basic FM synthesizer for SIREN.}
    \label{list:fm}
\end{listing}
The FM synthesis module defined in listing \ref{list:fm} follows the parameters set by John Chowning in his seminal article on the implementation of FM synthesis \cite{Chowning_1973}. Thus, in lines 1-10 in listing \ref{list:fm}, we can see the parameters \verb|p3| through \verb|p8|, representing duration, amplitude, carrier frequency, modulating frequency, and modulation indices one and two respectively. As seen in listing \ref{list:fm} lines 13 to 24 show the actual routing of the audio graph in Web Audio. The most interesting part of the code is the \verb|updates| property of the returned object. Starting on line 27, the \verb|updates| property defines a map that transforms changes in the parameter to changes in the audio node parameters. For example, the property on line 34 shows how the modulator amplitude is affected by changes in either \verb|p6| (modulating frequency), \verb|p7| (modulation index one), or \verb|p3| (duration). Thus, when any of these parameters are changed via the data, the update will trigger. The \verb|start| and \verb|stop| functions start and stop the various audio nodes.
\subsubsection{Working Example: Formant Synthesis in SIREN}
% \begin{figure}[!h]
%     \centering
%     \begin{subfigure}[!h]{0.45\textwidth}
%         \centering
%         \includegraphics[width=\textwidth]{assets/formant-parameters.png}
%         \caption{Formant synthesizer parameters.}
%         \label{fig:formant parameters}
%     \end{subfigure}
% \end{figure}
% \begin{figure}[!h]
%     \centering
%     \ContinuedFloat
%     \begin{subfigure}[!h]{0.45\textwidth}
%         \centering
%         \includegraphics[width=\textwidth]{assets/formant-worklets.png}
%         \caption{The impulse train audio worklet for the formant synthesizer.}
%         \label{fig:formant worklets}
%     \end{subfigure}
% \end{figure}
% \begin{figure}[!h]
%     \centering
%     \ContinuedFloat
%     \begin{subfigure}[!h]{0.45\textwidth}
%         \centering
%         \includegraphics[width=\textwidth]{assets/formant-default.png}
%         \caption{The main synthesizer code for the formant synthesizer.}
%         \label{fig:formant default}
%     \end{subfigure}
%     \caption{Basic formant synthesizer for SIREN.}
%     \label{fig:formant}
% \end{figure}
\begin{listing}[!h]
    \centering
\begin{minted}{js}
export const parameters = {
  timbral: { Frequency: 'quantitative', Vowel: 'nominal' },
  time: ['Time']
};

export const worklets = {
  impulse: class ImpulseTrainProcessor extends AudioWorkletProcessor {
    static get parameterDescriptors() { return [{ name: 'frequency', defaultValue: 200 }]; }

    constructor(options) {
      super(options);
      this.phase = 0;
    }

    process(inputs, outputs, parameters) {
      const output = outputs[0];
      const angularVelocity = 2 * Math.PI * parameters.frequency[0] / 44100;

      for (let channel = 0; channel < output.length; ++channel) {
        for (let i = 0; i < output[channel].length; ++i) {
          this.phase += angularVelocity;
          if (this.phase < Math.PI) output[channel][i] = 0;
          else {
            this.phase -= 2 * Math.PI;
            output[channel][i] = 1;
          }
        }
      }

      return true;
    }
  }
};

const formants = Object.freeze({
  a: { f1: 650, f2: 1080, f3: 2650 },
  e: { f1: 400, f2: 1700, f3: 2600 },
  i: { f1: 290, f2: 1870, f3: 2800 }
});

export default async (context, destination) => {
  await context.audioWorklet.addModule('impulse');

  const i = new AudioWorkletNode(context, 'impulse');
  const ig = new GainNode(context);
  const t = new IIRFilterNode(context, { feedback: [1], feedforward: [1, 0, -1] });
  const t2 = new IIRFilterNode(context, { feedback: [1], feedforward: [1, 0, 1] });
  const p = new IIRFilterNode(context, { feedback: [1, -0.99], feedforward: [0.01] });
  const f1 = new BiquadFilterNode(context, { type: 'bandpass' });
  const f2 = new BiquadFilterNode(context, { type: 'bandpass' });
  const f3 = new BiquadFilterNode(context, { type: 'bandpass' });
  const f1g = new GainNode(context);
  const f2g = new GainNode(context);
  const f3g = new GainNode(context);
  i.connect(ig).connect(t).connect(t2).connect(p)
    .connect(f1).connect(f1g).connect(destination);
  p.connect(f2).connect(f2g).connect(destination);
  p.connect(f3).connect(f3g).connect(destination);

  ig.gain.value = 500;

  f1g.gain.value = 3; f2g.gain.value = 0.8 * 3; f3g.gain.value = 0.6 * 3;
  
  f1.Q.value = 100; f2.Q.value = 100; f3.Q.value = 100;

  return ({
    updates: new Map()
      .set(['Frequency', 'Time'], (frequency = 200, time = 0) =>
        i.parameters.get('frequency').setValueAtTime(frequency, time)
      )
      .set(['Vowel', 'Time'], (vowel = 'i', time = 0) => {
        f1.frequency.setValueAtTime(formants[vowel].f1, time);
        f2.frequency.setValueAtTime(formants[vowel].f2, time);
        f3.frequency.setValueAtTime(formants[vowel].f3, time);
      }),
    start: () => {},
    stop: () => {}
  });
};
\end{minted}
    \caption{Basic formant synthesizer for SIREN.}
    \label{list:formant}
\end{listing}
The code in listing \ref{list:formant} follows the conventions set above and defines a basic formant synthesizer for parameter mapping sonification in SIREN. This example also showcases not only the flexibility, extensibility, and customizability of SIREN's parameter mapping paradigm, but also the full extent of available options of tools that SIREN provides for creating its synthesizer modules.

By examining lines 1-4 in listing \ref{list:formant}, we can first see the two timbral parameters that this basic formant synthesizer provides for parameter mapping sonification: \verb|Frequency| and \verb|Vowel|, along with the simple \verb|Time| temporal parameter that ensures that the changes to the output occur in time. Noticing on line 38 that the formant frequencies are already defined, the vowel parameter takes on the nominal data type with defined behavior for values of ``0,'' ``1,'' and ``2'' that correspond to the vowels ``a,'' ``e,'' and ``i'' respectively. The \verb|Frequency| parameter, on the other hand, is quantitative, since the pitch at which the vowels are uttered can be any numerical value.

Further along, we see that the optional \verb|worklets| export is also defined in listing \ref{list:formant} on line 6. Since the Web Audio API does not have an audio node for generating impulse trains, we must define our own, the easiest method of implementation being through \verb|AudioWorklet|s. Thus, we define a worklet with the name and module URL of \verb|impulse|, with the implementation of the impulse train generator as the value. Taking a peek at the parameters, we can see that the impulse train generator has a frequency parameter, which dictates how fast impulses should be generated. 

Finally, we reach the default export in listing \ref{fig:formant} on line 43, which defines the actual synthesizer. Lines 44 to 74 set up the audio graph as well as populate the audio nodes with initial values. In the returned object, we see that the three properties are defined. \verb|updates| creates a map with two update functions: one for \verb|Frequency| and \verb|Time|, and the other for \verb|Vowel| and \verb|Time|. The first property updates the frequency parameter of the impulse node, and the latter introduces a switch statement that adjusts the formant frequencies based on the nominal value of the parameter controlled by the data. Then, the start function is empty as nothing needs to be started, and the only thing that needs to stop when \verb|stop| is called is to suspend the context.
\subsubsection{Working Example: Granular Synthesis in SIREN}
% \begin{figure}[!h]
%     \centering
%     \begin{subfigure}[!h]{0.45\textwidth}
%         \centering
%         \includegraphics[width=\textwidth]{assets/granular-parameters.png}
%         \caption{Granular synthesizer parameters.}
%         \label{fig:granular parameters}
%     \end{subfigure}
% \end{figure}
% \begin{figure}[!h]
%     \centering
%     \ContinuedFloat
%     \begin{subfigure}[!h]{0.45\textwidth}
%         \centering
%         \includegraphics[width=\textwidth]{assets/granular-default.png}
%         \caption{The main synthesizer code for the granular synthesizer.}
%         \label{fig:granular default}
%     \end{subfigure}
%     \caption{Basic granular synthesizer for SIREN.}
%     \label{fig:granular}
% \end{figure}
\begin{listing}[!h]
    \centering
\begin{minted}{js}
export const parameters = {
  timbral: {
    Buffer: 'nominal',
    Rate: 'quantitative',
    Position: 'quantitative',
    Gain: 'quantitative'
  },
  time: ['Duration', 'Time']
};

const bufferize = async (context, url) =>
  await context.decodeAudioData(
    await (
      await fetch(url, { cache: 'force-cache' })
    ).arrayBuffer()
  );

export default async (context, destination) => {
  const buffers = {
    buffer1: await bufferize(context, 'url'),
    buffer2: await bufferize(context, 'url')
  };

  return ({
    updates: new Map()
      .set(['Buffer',
            'Rate',
            'Position',
            'Gain',
            'Time',
            'Duration'
           ],
        (Buffer = 'buffer1',
         Rate = 1,
         Position = 0,
         Gain = 1,
         Time = 0,
         Duration = 0.25
        ) => {
          const source = new AudioBufferSourceNode(context);
          const amp = new GainNode(context);
          source.connect(amp).connect(destination);
          source.buffer = buffers[Buffer];
          amp.gain.value = Gain;
          source.playbackRate.value = Rate;
          source.start(
            Time,
            buffers[Buffer].duration * Position,
            Duration
          );
        }),
    start: () => {},
    stop: () => {}
  });
};
\end{minted}
    \caption{Basic granular synthesizer for SIREN.}
    \label{list:granular}
\end{listing}
This granular synthesizer demonstrates another important aspect of SIREN's sound synthesizer module paradigm: the ability to request outside resources. After defining a few important parameters on lines 1-9 in listing \ref{list:granular}, lines 19-22 initialize two audio buffers, and the rest of the audio graph is created in the update map. The smallest unit in a granular synthesizer is a grain, so there is only one update function, as all parameters of a grain should be updated at the same time as the grain is queued for playback. In the example, the implemented parameters include buffer, playback rate, grain position, gain, grain duration, and time. The only update concerns all of the parameters; as reasoned above, since the grain is the smallest unit of sound, this sets all of the properties of the grain whenever a new grain is queued. Lines 40-50 in listing \ref{list:granular} instantiate the audio graph. Knowing the values of the various characteristics of the grain, the code instantiates a new \verb|AudioBufferSourceNode| to set the buffer (line 43), rate (line 45), position (line 48), duration (line 49), and start (line 47), and \verb|GainNode| to set the gain (line 44). Line 42 connects the subgraph to the overall larger audio graph. Since updates to multiple parameters that occur at the same time only call the update function once, simultaneous sonification of multiple parameters are correctly resolved.
\section{Future Work}
While the examples in this paper have demonstrated a wide breadth of sound synthesis paradigms that SIREN can support easily, there still can be more work done to incorporate more types of sound synthesis into SIREN to truly demonstrate its capabilities as a general purpose data sonification interface.

Another powerful general purpose sonification interface is SonART, which uses the Synthesis ToolKit to provide the sonification parameters. With the development of Emscripten and WebAssembly, future work could involve the compilation of STK into WebAssembly, which would allow SIREN to also use STK for parameter mapping sonification.

Apart from unlocking the potential of STK on the web, WebAssembly also affords a lot of other technology, such as WebChucK, Faust RNBO, Csound\footnote{https://github.com/csound/csound/tree/master/wasm/browser}, WebPd\footnote{https://github.com/sebpiq/WebPd} \cite{nime2023_28, Letz_Orlarey_Fober_2018}. Their availability on the web through WebAssembly makes it possible for SIREN to integrate with these technologies to create synthesizers, expanding the cross-sectional area of what currently is the bottleneck in the current workflow of creating sonifications in SIREN. Web Audio Modules is also another standard for writing plugins in Web Audio, which could provide yet another avenue for simplifying the process of creating sounds and parameters for SIREN.

Beyond parameter mapping sonification, another sonification method used by a lot of other sonification interfaces is audification \cite{Pauletto_Hunt_2004, Dombois_2008, Phillips_Cabrera_2019}. While we expect audification to be possible in SIREN in its current implementation, future work would entail creating an audification synthesizer.

Most importantly, an ancillary project to SIREN would entail the creation of an open-sourced synthesizer repository where users can submit synthesizers for others to use. Since the current limiting factor of SIREN in achieving more general purpose usage is the methods of sonification, a crowd-sourced effort in developing more synthesizer implementations using the JavaScript and the Web Audio API can kickstart a positive feedback loop.
\section{Conclusion}
SIREN works to thread the needle of raising the ceiling while maintaining as low a threshold as possible as a general purpose data sonification interface. However, through SIREN's design, vast possibilities for aesthetic sonifications of all kinds are possible in SIREN. This paper demonstrated three examples of various complexity, and future work includes the further development of synthesizers that can better cover the infinite possibilities of synthesized sounds and its parameters.
%\end{document}  % This is where a 'short' article might terminate

%ACKNOWLEDGMENTS are optional
% \section{Acknowledgments}
% This section is optional; it is a location for you
% to acknowledge grants, funding, editing assistance and
% what have you.  In the present case, for example, the
% authors would like to thank Gerald Murray of ACM for
% his help in codifying this \textit{Author's Guide}
% and the \textbf{.cls} and \textbf{.tex} files that it describes.

\bibliographystyle{abbrv}
\bibliography{sigproc}  % sigproc.bib is the name of the Bibliography in this case

\begin{thebibliography}{10}

\bibitem{Hermann_Hunt_Neuhoff_2011}
{\em The sonification handbook}.
\newblock Logos Verlag, Berlin, 2011.

\bibitem{Barrass_Kramer_1999}
S.~Barrass and G.~Kramer.
\newblock Using sonification.
\newblock {\em Multimedia Systems}, 7(1):23–31, Jan. 1999.

\bibitem{Ben-Tal_Berger_Cook_Daniels_Scavone_2002}
O.~Ben-Tal, J.~Berger, B.~Cook, M.~Daniels, and G.~Scavone.
\newblock Sonart: The sonification application research toolbox.
\newblock July 2002.

\bibitem{Bruce_Palmer_2005}
J.~W. Bruce and N.~T. Palmer.
\newblock Sift: Sonification integrable flexible toolkit.
\newblock July 2005.

\bibitem{Candey_Schertenleib_Diaz-Merced_2006}
R.~M. Candey, A.~M. Schertenleib, and W.~L. Diaz-Merced.
\newblock Xsonify sonification tool for space physics.
\newblock June 2006.

\bibitem{Cantrell_Walker_Moseng_2021}
S.~J. Cantrell, B.~N. Walker, and {\O}.~Moseng.
\newblock Highcharts sonification studio: An online, open-source, extensible, and accessible data sonification tool.
\newblock In {\em Proceedings of the 26th International Conference on Auditory Display (ICAD 2021)}, page 210–216, Virtual Conference, June 2021. International Community for Auditory Display.

\bibitem{Cherston_Paradiso_2017}
J.~Cherston and J.~A. Paradiso.
\newblock Rotator: Flexible distribution of data across sensory channels.
\newblock {\em other univ website}, June 2017.

\bibitem{Chowning_1973}
J.~M. Chowning.
\newblock The synthesis of complex audio spectra by means of frequency modulation.
\newblock {\em Journal of the Audio Engineering Society}, 21(7):526–534, Sept. 1973.

\bibitem{Cook_Scavone_1999}
P.~R. Cook and G.~P. Scavone.
\newblock The synthesis toolkit (stk).
\newblock In {\em Proceedings of the 1999 International Computer Music Conference}, Beijing, China, 1999.

\bibitem{Davison_Walker_2007}
B.~K. Davison and B.~N. Walker.
\newblock Sonification sandbox reconstruction: Software standard for auditory graphs.
\newblock June 2007.

\bibitem{Dombois_2008}
F.~Dombois.
\newblock Sonifyer a concept, a software, a platform.
\newblock June 2008.

\bibitem{Garcia_Diaz-Merced_Casado_Cancio_2019}
B.~Garcia, W.~Diaz-Merced, J.~Casado, and A.~Cancio.
\newblock Evolving from xsonify: a new digital platform for sonorization.
\newblock {\em EPJ Web of Conferences}, 200:01013, 2019.

\bibitem{Joseph_Lodha_2002}
A.~J. Joseph and S.~K. Lodha.
\newblock Musart: Musical audio transfer function real-time toolkit.
\newblock July 2002.

\bibitem{Letz_Orlarey_Fober_2018}
S.~Letz, Y.~Orlarey, and D.~Fober.
\newblock Faust domain specific audio dsp language compiled to webassembly.
\newblock In {\em Companion Proceedings of the The Web Conference 2018}, WWW ’18, page 701–709, Republic and Canton of Geneva, CHE, Apr. 2018. International World Wide Web Conferences Steering Committee.

\bibitem{Lindetorp_Falkenberg_2020}
H.~Lindetorp and K.~Falkenberg.
\newblock Webaudioxml: Proposing a new standard for structuring web audio.
\newblock June 2020.

\bibitem{Lindetorp_Falkenberg_2021}
H.~Lindetorp and K.~Falkenberg.
\newblock Sonification for everyone everywhere - evaluating the webaudioxml sonification toolkit for browsers.
\newblock In {\em Proceedings of the 26th International Conference on Auditory Display (ICAD 2021)}, page 15–21, Virtual Conference, June 2021. International Community for Auditory Display.

\bibitem{Lodha_Beahan_Heppe_Joseph_Zane-Ulman_1997}
S.~K. Lodha, J.~Beahan, T.~Heppe, A.~Joseph, and B.~Zane-Ulman.
\newblock Muse: A musical data sonification toolkit.
\newblock Nov. 1997.

\bibitem{Myers_Hudson_Pausch_2000}
B.~Myers, S.~E. Hudson, and R.~Pausch.
\newblock Past, present, and future of user interface software tools.
\newblock {\em ACM Transactions on Computer-Human Interaction}, 7(1):3–28, Mar. 2000.

\bibitem{Pauletto_Hunt_2004}
S.~Pauletto and A.~Hunt.
\newblock A toolkit for interactive sonification.
\newblock 2004.

\bibitem{Peng_Choi_2021}
T.~Peng and H.~Choi.
\newblock Siren: A case study in web audio based sonification.
\newblock In {\em Proceedings of the 26th International Conference on Auditory Display (ICAD 2021)}, page 126–130, Virtual Conference, June 2021. International Community for Auditory Display.

\bibitem{Peng_Choi_Berger_2023}
T.~Peng, H.~Choi, and J.~Berger.
\newblock Siren: Creative and extensible sonification on the web.
\newblock In {\em Proceedings of the 28th International Conference on Auditory Display (ICAD2023)}, page 78–84, Linköping University, June 2023. International Community for Auditory Display.

\bibitem{Phillips_Cabrera_2019}
S.~Phillips and A.~Cabrera.
\newblock Sonification workstation.
\newblock June 2019.

\bibitem{Stevens_1946}
S.~S. Stevens.
\newblock On the theory of scales of measurement.
\newblock {\em Science}, 103(2684):677–680, June 1946.

\bibitem{Walker_Cothran_2003}
B.~N. Walker and J.~T. Cothran.
\newblock Sonification sandbox: A graphical toolkit for auditory graphs.
\newblock July 2003.

\end{thebibliography}
\end{sloppypar}
\end{document}